# A Swarming Approach to Optimize the One-hop Delay in Smart Driving Inter-platoon Communications

**Qiong Wu[1,2,3], Shuzhen Nie[1], Pingyi Fan[3], Hanxu Liu[1], Fan Qiang[4] and Zhengquan Li[1,2]**
[1]Jiangsu Provincial Engineering Laboratory of Pattern Recognition and Computational Intelligence, Jiangnan University, Wuxi, 214122, China
[2]National Mobile Communications Research Laboratory, Southeast University, Nanjing, 210096, China
[3]Department of Electronic Engineering, Tsinghua University, Beijing 100084, China
[4]Advanced Networking Lab., Department of Electrical and Computer Engineering, New Jersey Institute of Technology, Newark, NJ 07102, USA

Corresponding author: Pingyi Fan and Zhengquan Li.

**ABSTRACT** In this paper, we propose a swarming approach and optimize the one-hop delay for inter-platoon communications through adjusting the minimum contention window size of each backbone vehicle in two steps. In the first step, we first set a small enough average one-hop delay as the initial optimization goal and then propose a swarming approach to find a minimum average one-hop delay for inter-platoon communications through adjusting the minimum contention window of each backbone vehicle iteratively. In the second step, we first set the minimum average one-hop delay found in the first step as the initial optimization goal and then adopt the swarming approach again to get the one-hop delay of each backbone vehicle balance to the minimum average one-hop delay. The optimal minimum contention window sizes that get the one-hop delay of each backbone vehicle balance to the minimum average one-hop delay are obtained after the second step. The simulation results indicate that the one-hop delay is optimized and the other performance metrics including end-to-end delay, one-hop throughput and transmission probability are presented by using the optimal minimum contention window sizes.

## I. INTRODUCTION

Autonomous driving has been a promising technology in recent years. It can improve road safety by sensing the road environment, including vehicles, pedestrians and obstacles, through communications between vehicles and vehicles, i.e., vehicle-to-vehicle (V2V) communications, and communications between vehicles and infrastructures, i.e., vehicle-to-infrastructure (V2I) communications [1]. It can also satisfy users' entertainment demand by liberating them from a lot of driving time, enabling them to drink coffee, deal with their business and so on [2]. Due to these benefits, many automobile manufacturers and academic institutions have focused on autonomous driving technology. For example, in 2010, Google X Lab developed the first driverless car and tested it successfully in California [3]. In 2014, Baidu, teaming up with Bavarian Motor Work (BMW), started up an autonomous driving research project, and in 2015, tested the developed autonomous vehicles on complicated roads in Beijing and Shanghai [4].

Platooning is an important management strategy for autonomous vehicles. With platooning strategy, autonomous vehicles periodically transmit a packet with vehicles' kinematics information including velocity and acceleration rate through V2V communications to keep a constant speed and a small constant distance one after another on the common lane [5]. In this case, autonomous vehicles follow one after another in a queue and organize themselves into a set called a platoon. Each platoon includes a leader vehicle, tail vehicle and some member vehicles. In a platoon, the leader vehicle is the first vehicle, which will control the speed and driving direction of the platoon. The tail vehicle is the last vehicle within the platoon. The member vehicles are these with middle places in the queue. The leader vehicle and tail vehicle are called backbone vehicles and the member vehicles are called non-backbone vehicles. It has been pointed out that platooning strategy can reduce traffic congestion, conserve energy consumption, improve traffic safety and facilitate the management of autonomous vehicles [6].

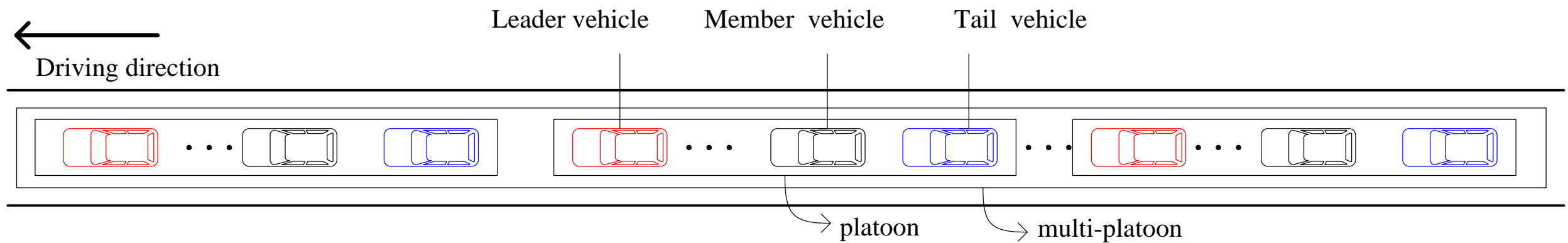

FIGURE 1. **A multi-platoon scenario**

However, with the number increment of vehicles in a platoon, the vehicles in a platoon are difficult to be managed simultaneously. To solve this problem, vehicles are organized into a multi-platoon instead of a single platoon [7]. A multi-platoon consists of a few platoons, as shown in Figure 1. In each platoon, the backbone vehicles are equipped with two transceivers [8]. One transceiver is used for intra-platoon communications, i.e., backbone vehicles communicate with the member vehicles in a same platoon. Another transceiver is used for inter-platoon communications, i.e., backbone vehicles communicate with the other backbone vehicles. For a multi-platoon, the backbone vehicles in each platoon periodically transmit a packet with vehicles' kinematics information through inter-platoon communications, and then the backbone vehicles in a platoon forwards this packet to the member vehicles of the platoon through intra-platoon communications to keep a platoon formation.

For inter-platoon communications, backbone vehicles adopt the IEEE 802.11 distributed coordination function (DCF) mechanism to access a control channel (CCH) and communicate with each other through multi-hop communications [8]. In this case, there are several major factors that may affect the one-hop delay.

(1) The first one is the minimum contention window size defined in IEEE 802.11 DCF mechanism. In fact, the IEEE 802.11 DCF mechanism uses a carrier sense multiple access with collision avoidance (CSMA/CA) mechanism to access a channel. It adopts a binary exponential back-off rule to reduce collision. According to the IEEE 802.11 standard [9], the minimum contention window size is 64, which is too large. In this case, a vehicle would spend a lot of time during a back-off procedure before transmitting a packet, therefore, the one-hop delay is relatively long, may not be suitable for V2V communications.

(2) The second factor is the hidden terminal problem. For inter-platoon communications, if two backbone vehicles which are not in the communication range of each other, their transmitted a packet may arrive at a third backbone vehicle causing a collision refer to as at the third backbone vehicle. This is the hidden terminal problem. One of the two backbone vehicles is the hidden terminal of another backbone vehicle. The hidden terminal problem would cause the two backbone vehicles retransmit the packet, resulting in the one-hop delay increased. In a multi-platoon, the number of hidden terminals of each backbone vehicle is different, so the one-hop delay of these backbone vehicles would be unbalanced.

(3) The third one is the number of neighbors. The number of neighbors may be different for different backbone vehicles. The number of neighbors for the first and last backbone vehicle is 1 and the number of neighbors is 2 for the other backbone vehicles. When a backbone vehicle is transmitting a packet, if more than one neighbor is transmitting other packets at the same time, there would be a collision. In this case, the backbone vehicle would retransmit the packet and increase the one-hop delay. For different backbone vehicles, if the number of neighbors is different, the collision probability would be different and thus cause an unbalanced one-hop delay.

The long and unbalanced one-hop delay would affect the end-end delay from each source backbone vehicle to its destination backbone vehicle. If the one-hop delay of a backbone vehicle is long, the end-to-end delay from this backbone vehicle to its destination backbone vehicle would also be long. This would cause that the destination backbone vehicle cannot receive vehicles' kinematics information timely through inter-platoon communications. It would cause that the destination backbone vehicle cannot forward these information to its member vehicles in time and thus the member vehicles cannot know some emergency changes of vehicles' kinematics within a limited time and take prompt action accordingly to keep a platoon formation. Therefore, the one-hop delay is an important performance metric for inter-platoon communications. To the best of our knowledge, there is no research on designing an approach to optimize the one-hop delay for inter-platoon communications. That motivates us to conduct this work.

In this paper, we propose a swarming approach and optimize the one-hop delay for inter-platoon communications through adjusting the minimum contention window size of each backbone vehicle in two steps. In the first step, we first set a small enough average one-hop delay of backbone vehicles as the initial optimization goal and then propose a swarming approach to find a minimum average one-hop delay for inter-platoon communications through adjusting the minimum contention window of each backbone vehicle iteratively. In the second step, we first set the minimum average one-hop delay found in the first step as the initial optimization goal and then adopt the swarming approach again to get the one-hop delay of each backbone vehicle balance to the minimum average one-hop delay. The optimal minimum contention window sizes that get the one-hop delay of each backbone vehicle balance to the minimum average one-hop delay are obtained after the second step. The simulation results indicate that the one-hop delay is optimized and the other performance metrics including end-to-end delay, one-hop throughput and transmission probability are discussed by using the optimal minimum contention window sizes.

The rest of this paper is organized as follows. Second II reviews the relevant recent works. The multi-platoon scenario is introduced in section III. The procedure that optimizes the one-hop delay is described in section IV. In section V, we evaluate the network performance, and finally, we conclude this paper in section VI.

## II. RELATED WORK

Many works developed some models to describe the performance of VANETs (vehicular ad hoc networks) or designed novel access schemes to improve the performance of VANETs [8,10-16]. However, these works did not consider the platooning strategy.

Some works mainly focused on designing approaches to enhance the access performance of platoon communications including reliability of its data transmission, packet delay and network connectivity.

In [17], Shao et al. used a multi-priority Markov chain model to explore the relationship between the network connectivity probability and the system throughput under various traffic densities for platoon-based VANETs. It has shown that the system throughput would increase with the increment of network connectivity probability in a certain range.

In [18], Ye et al. proposed a medium access control (MAC) protocol based on the IEEE 802.11 DCF mechanism. This protocol tried to guarantee the emergency warning message (EWM) to be transmitted successfully within a low delay, and thus vehicles in a platoon can receive the messages and react timely to avoid some rear-end collisions on a highway.

In [19], Bohm and Kunert proposed a slot-based retransmission approach for broadcasting. This approach set a retransmission opportunity for the control message according to the data age of the last received data. Moreover, it set priority to resend the control message. This approach can cut down the danger degree efficiently.

In [20], Guo et al. proposed a risk-aware MAC protocol. It can make a good trade-off between effectiveness and fairness for delivering messages, such as the warning messages and beacon messages. Moreover, it provided a way for real-time and reliable security messages among vehicles in a platoon.

In [21], Campolo et al. designed a full-duplex (FD) MAC protocol based on the IEEE 802.11 standard in a multi-channel vehicular network. This work has shown that the FD technology enhanced the collision-detection mechanism, boosted the channel capacity and brought higher reliability for broadcasting safety messages. In addition, it enhanced the platoon control.

In [22], Ali Balador et al. proposed a reliable token-passing access approach to maintain the flexibility of the IEEE 802.11p access mechanism and overcome its drawback on messages' reliability. This approach can reduce the data age and increase the transmission reliability.

In [23], Su and Zhang proposed an approach which integrated a clustering contention-free MAC mechanism and the contention-based IEEE 802.11 MAC mechanism. In this approach, the cluster-head vehicles adopted it to guarantee the safety messages transmitted on time. In addition, this approach could obtain an optimal contention-window size to get the best balance between the transmission delay and the successful transmission rate.

As mentioned above, some approaches have been proposed to reduce delay and achieve a reliable transmission. However, the approaches mentioned above did not consider a multi-platoon scenario and the optimization of the one-hop delay for each backbone vehicle in inter-platoon communication. A long and unbalanced one-hop delay of each backbone vehicle for inter-platoon communication may affect the end-end delay and thus may pose a potential threat to safe driving, especially for a highway scenario. To solve this problem, we propose a swarming approach to reduce and balance the one-hop delay of each backbone vehicle for inter-platoon communications.

## III. SYSTEM MODEL

For a multi-platoon, the backbone vehicles in each platoon periodically transmit a packet with vehicles' kinematics information through inter-platoon communications, and then a backbone vehicle in a platoon forwards this packet to its platoon members through intra-platoon communications to keep a platoon formation. In this paper, we focus on the inter-platoon communications. In this section, we introduce the inter-platoon communication model.

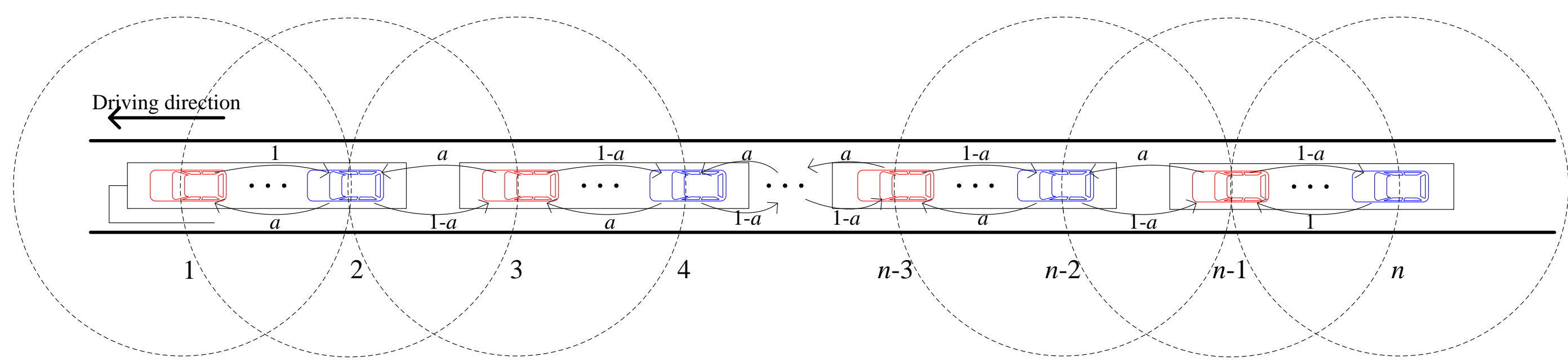

FIGURE 2. **Inter-platoon communications model**

Considering a multi-platoon which is moving in a highway. Let $n$ be the number of backbone vehicles in the multi-platoon. We label these backbone vehicles with $1,2,\ldots,n\text{-}1,n$. For the inter-platoon communications, the backbone vehicles use the same transceiver to communicate with each other through multi-hop communications. The data traffic is saturated, i.e., a vehicle always has packets to transmit. Let $a$ be the probability that vehicle $i$ is the destination of vehicle $i$+1's packet, where $1\leq i<n\text{-}1$. The probability that vehicle $i$+2 is the destination of vehicle $i$+1's packet is 1-$a$. The inter-platoon communications model is illustrated in Figure. 2.

For the inter-platoon communications, backbone vehicles adopt the IEEE 802.11 DCF mechanism to access a CCH. When a backbone vehicle has a packet to transmit, it first selects an integer from $[0,W_k\text{-}1]$ randomly as the value of the back-off counter, here $k$ denotes the number that the packet is retransmitted and $W_k$ is the minimum contention window size when the packet is retransmitted for $k$ times. If the channel is detected idle, the value of the back-off counter would be decremented by one. Otherwise, the value of the back-off counter would be frozen until the channel is detected idle continuously during a distributed inter-frame spacing (DIFS) period. If the value of the back-off counter is decremented to 0, the backbone vehicle would transmit the packet. If the backbone vehicle does not receive an acknowledge (ACK) message after a short inter-frame space (SIFS) period, it would select a value from $[0,W_{k+1}\text{-}1]$ randomly as the value of the back-off counter and initial a new back-off procedure, here $W_{k+1}=2W_k$. If the number of retransmission reaches a retransmission limit, the backbone vehicles would drop this packet.

## IV. SWARMING APPROACH DESCRIPTION

In this section, we propose a swarming approach to optimize the one-hop delay for inter-platoon communications in two steps, i.e., step A and step B. In step A, a swarming approach is proposed to find a minimum average one-hop delay of inter-platoon communications through adjusting the minimum contention window of each backbone vehicle iteratively. In step B, the proposed swarming approach is used again to get the one-hop delay of each backbone vehicle balance to the minimum average one-hop delay found in step A. After the second step, the optimal minimum contention window sizes with which the optimal one-hop delay of inter-platoon communications are obtained.

### *A. Minimum Average One-hop Delay*

In step A, a swarming approach is proposed to find a minimum average one-hop delay of inter-platoon communications iteratively. In each iteration, the minimum contention window sizes of backbone vehicles are adjusted to enable the one-hop delays of backbone vehicles get close to an optimization objective. The closeness between the one-hop delays of backbone vehicles and the optimization objective is evaluated by an optimization function. Before the swarming approach is initialed, an optimization objective and an objective function are determined. The optimization objective is a small enough average one-hop delay $D_{avg}$. The value of $D_{avg}$ is chosen according to plenty of simulation experiments. The simulation scenario is described in section III. Each simulation experiment is conducted with different minimum contention window sizes. The value of $D_{avg}$ is lower than the smallest average one-hop delay obtained by the simulation experiments. The variance between the one-hop delays of backbone vehicles and $D_{avg}$ can be used to evaluate the closeness between them. Hence the objective function is defined as the variance between the one-hop delays of backbone vehicles and $D_{avg}$.

Next, we will introduce the swarming approach in detail. The procedure of the swarming approach is divided into 3 sub-procedure, i.e., initialization procedure, iteration procedure and output procedure.

#### *a) Initialization Procedure*

In the initialization procedure, $m$ combinations of minimum contention window sizes are initialized. Each combination includes the minimum contention window sizes of $n$ backbone vehicles. Each minimum contention window size is an integer randomly selected from [1,64]. Here $m$ is an integer that cannot be too large or too small. If it is too large, the complexity of the swarming approach would be increased. If it is too small, the minimum average one-hop delay may not be found.

***b) Iteration Procedure***

In the iteration procedure, each combination including $n$ minimum contention window sizes are updated iteratively to enable the one-hop delays of backbone vehicles get close to $D_{avg}$. In each iteration, there are a global optimal solution for all combinations and an individual optimal solution for each combination. In the $t^{th}$ ($t$>1) iteration, the global optimal solution $g(t)$ is the combination that minimizes the value of the objective function among all the updated combinations until the $(t\text{-}1)^{th}$ iteration. The individual optimal solution of combination $j$ $p_j(t)$ is the combination that minimizes the value of the objective function among all the updated combinations of combination $j$ until the $(t\text{-}1)^{th}$ iteration. In the first iteration ($t$=1), $g(1)$ is the combination that minimizes the value of the objective function among all the initialed combination and $p_j(1)$ is the initial combination $j$.

For the first iteration, the input is the initial $m$ combinations. Next the one-hop delays corresponding to each combination are obtained through simulation experiments and the value of the objective function corresponding to each combination is calculated. Then the combination corresponding to the minimum value of the objective function is selected as $g(1)$ and combination $j$ is selected as $p_j(1)$. If $g(1)$ is smaller than a threshold, the iteration procedure would stop and the one-hop delays are the output. Otherwise, each combination is updated according to $g(1)$, $p_j(1)$ and some parameters at the end of the first iteration. For the $t^{th}$ ($t$>1) iteration, the input is the $m$ combinations updated at the end of $(t\text{-}1)^{th}$ iteration. Next the one-hop delays and the value of the objective function corresponding to each combination are calculated like the first iteration. Then $g(t)$ is selected through comparing the minimum value of the objective function and the value of the objective function corresponding to $g(t\text{-}1)$, the combination corresponding to the minimum value of them is selected as $g(t)$. Similarly, $p_j(t)$ is selected through comparing the value of the objective function corresponding to combination $j$ and the value of the objective function corresponding to $p_j(t\text{-}1)$, the combination corresponding to the minimum value of them is selected as $p_j(t)$. If $g(t)$ is smaller than a threshold or the number of iterations reaches the predefined maximum value, the iteration would stop and the corresponding one-hop delays are the output. Otherwise, each combination is updated according to $g(t)$, $p_j(t)$ and some parameters at the end of the $t^{th}$ iteration. Next, we introduce each iteration procedure in detail.

For the $t^{th}$ iteration, the input is the initial $m$ combinations ($t$=1) or the $m$ combinations updated at the end of $(t\text{-}1)^{th}$ iteration ($t$>1). Let $cw_{ij}(t)$ be the minimum contention window size of backbone vehicle $i$ in combination $j$. The one-hop delay of each backbone vehicle with the minimum contention window size in a combination is measured through a simulation experiment. The simulation scenario is described in section III. In the simulation experiment, the backbone vehicle $i$ in combination $j$ transmits packets with the minimum contention window size $cw_{ij}(t)$. The one-hop delay of the backbone vehicle $i$ in combination $j$ is calculated by Eq. (1),

$$D_{ij}(t) = \frac{T_{ij}(t)}{x_{ij}(t)}, \tag{1}$$

where $D_{ij}(t)$ is the one-hop delay of backbone vehicle $i$ in combination $j$, $T_{ij}(t)$ is the duration that backbone vehicle $i$ in combination $j$ transmits packets, $x_{ij}(t)$ is the number of the packets transmitted successfully by backbone vehicle $i$ in combination $j$. Both $T_{ij}(t)$ and $x_{ij}(t)$ are measured in the simulation experiment.

Then the closeness between the one-hop delays of the backbone vehicles with the minimum contention windows in combination $j$ and $D_{avg}$ is measured through an objective function, i.e., the variance between them, which is given by Eq. (2),

$$f(CW_j(t)) = \sum_{i=1}^{n}[(D_{ij}(t) - D_{avg})]^2, \tag{2}$$

where $CW_j(t)$ denotes combination $j$ including the minimum contention window sizes of $n$ backbone vehicles, i.e., $CW_j(t)=\{cw_{1j}(t), cw_{2j}(t), \ldots, cw_{ij}(t), \ldots, cw_{nj}(t)\}$, $f(CW_j(t))$ denotes the closeness between the one-hop delays of backbone vehicles with the minimum contention windows in combination $j$ and $D_{avg}$.

Given $m$ combinations, $m$ values of the objective function are calculated according to Eq. (2). The combination that minimizes the value of the objective function is $gmin(t)$. Let $gmin_i(t)$ be the minimum contention window size of backbone vehicle $i$ in $gmin(t)$, $g_i(t)$ be the minimum contention window size of backbone vehicle $i$ in $g(t)$ and $p_{ij}(t)$ be the minimum contention window size of backbone vehicle $i$ in $p_j(t)$. When $t$>1, $g_i(t)$ is selected through comparing $f(gmin(t))$ with $f(g(t\text{-}1))$ and $p_{ij}(t)$ is selected through comparing $f(CW_j(t))$ with $f(p_j(t\text{-}1))$. The comparisons are shown as Eq. (3) and (4) respectively,

$$g_i(t) = \begin{cases} g_i(t-1), & f(g\min(t)) \geq f(g(t-1)), t>1 \\ g\min_i(t), & f(g\min(t)) < f(g(t-1)), t>1 \end{cases}, \tag{3}$$

$$p_{ij}(t)=\begin{cases} p_{ij}(t-1), & f(CW_j(t)) \geq f(p_j(t-1)), t>1 \\ cw_{ij}(t)\ , & f(CW_j(t)) < f(p_j(t-1)), t>1 \end{cases} . \tag{4}$$

When $t$=1, $g_i(t)$ is selected as $gmin_i(t)$ and $p_{ij}(t)$ is selected as $cw_{ij}(t)$.

The one-hop delays of backbone vehicles are judged whether they are close enough to $D_{avg}$. If $f(g(t))$ is smaller than a threshold value or the number of iterations reaches the predefined limit, the iteration procedure would be stopped and the one-hop delays of backbone vehicles with the minimum contention windows in $g(t)$ would be the output of the iteration procedure.

If $f(g(t))$ is not smaller than a threshold value and the number of iterations does not reach the limit, $cw_{ij}(t)$ would be updated according to $g(t)$, $p_j(t)$ and some parameters at the end of the $t^{th}$ iteration. When $t$>1, $cw_{ij}(t)$ is updated according to Eq. (5) and (6),

$$\begin{aligned}\Delta cw_{ij}(t) &= w\cdot\Delta cw_{ij}(t-1)+c_1\cdot r_1\cdot(g_i(t)-cw_{ij}(t)) \\ &\quad +c_2\cdot r_2\cdot(p_{ij}(t)-cw_{ij}(t))\end{aligned}\ , \tag{5}$$

$$cw_{ij}(t+1)=cw_{ij}(t)+\Delta cw_{ij}(t)\ , \tag{6}$$

where $c_1$ and $c_2$ are the learning coefficients and are positive constants, $r_1$ and $r_2$ are the random numbers in [0,1], $w$ is the inertia coefficient, $\Delta\ cw_{ij}(t)$ is the variation of $cw_{ij}(t)$. Here $\Delta\ cw_{ij}(t)$ cannot be larger than a constant $\Delta\ cw_{max}$. If $\Delta\ cw_{ij}(t)$ is larger than $\Delta\ cw_{max}$, it would keep $\Delta\ cw_{max}$. When $t$=1, $\Delta\ cw_{ij}(t)$ is a random value selected from [0,1] and $cw_{ij}(t+1)$ is updated according to Eq. (6).

#### *c) Output Procedure*

In the output procedure, the inputs are the output of iteration procedure, i.e., the one-hop delays of backbone vehicles with the minimum contention windows in $g(t)$. The minimum average one-hop delay is calculated by averaging the one-hop delays. Finally, the output of output procedure is the minimum average one-hop delay.

The pseudocode and the flow chart of the swarming approach are described as Algorithm 1 and Figure 3 respectively.

Algorothm 1. The pseudocode of the swarming approach

```
t=1;
//initialization procedure
Determine D_avg;
for j=1 to m (combinations)
        for i=1 to n (backbone vehicles) do
                initialize cw_ij(t);
        end
end
//iteration procedure
repeat
        calculate D_ij(t) through simulation;//Eq. (1)
        for j=1 to m do
        CW_j(t)={cw_1j(t), cw_2j(t), …, cw_ij(t), …, cw_nj(t)};
        calculate f(CW_j(t)) under D_avg;// Eq. (2)
        end
        gmin(t)=f^-1(min(f(CW_j(t))));
        if  t==1
              g(t)=gmin(t);
              p_j(t)= CW_j(t);
else
if  f(gmin(t))<f(g(t-1))
                        g(t)=gmin(t);
else
g(t)=g(t-1);
end
if  f(CW_j(t))<f(p_j(t-1))
                        p_j(t)= CW_j(t);
else
p_j(t)= p_j(t-1);
end
end
      if  f(g(t))<threshold or number of iterations == limit
              return  D_ij(t) with g(t)
              break;
      else
              for j=1 to m
```

for $i$=1 to $n$ do
compute $\triangle cw_{ij}(t)$ ;// Eq. (5)
update $cw_{ij}(t+1)$ ; //Eq. (6)
end
end
$t$++;
end
// output procedure
average $D_{ij}(t)$ ;
return the average result;

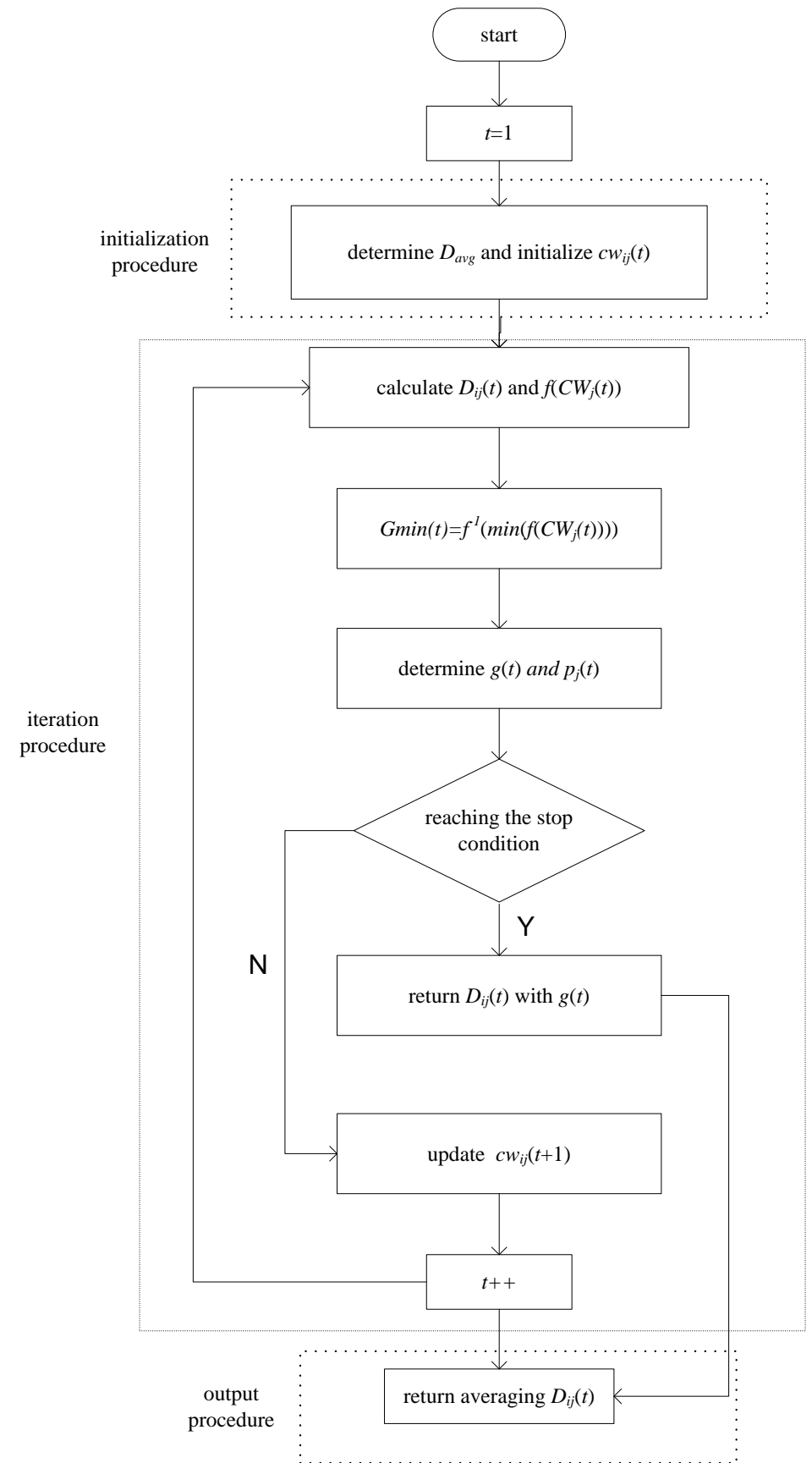

**FIGURE 3. The flow chart of the swarming approach**

### *B. Optimal Minimum Contention Windows*

After step A, we have obtained the minimum average one-hop delay. However, this does not guarantee that the one-hop delay of each backbone vehicle gets balance to the minimum average one-hop delay. In step B, the combination including the optimal minimum contention windows that gets the one-hop delay of each backbone vehicle balance to the minimum average one-hop delay is found.

In step B, the optimization objective is the minimum average one-hop delay, the objective function is the variance between the one-hop delays of backbone vehicles and the minimum average one-hop delay. The swarming approach proposed in step A is used again to get the one-hop delay of each backbone vehicle balance to the minimum average one-hop delay through adjusting the minimum contention window size. Finally, the output of the output procedure is the combination including the optimal minimum contention window sizes that gets the one-hop delay of each backbone vehicle balance to the minimum average one-hop delay.

## V. PERFORMANCE RESULTS

In this section, we compare the performance in terms of the transmission probability, one-hop delay, end-to-end delay, one-hop throughput and transmission probability of the inter-platoon communications under the optimal minimum contention window sizes and the standard minimum contention window size defined in the IEEE 802.11 DCF mechanism through simulation experiments. The simulation scenario is described in section III. Let $CW_{min}$ be the standard minimum contention window size defined in the IEEE 802.11 DCF mechanism, $E[L]$ be the size of each packet, $s$ be the duration of a slot, $M$ be the retransmission limit, $R$ be the channel bit rate, $p_e$ be transmission error probability caused by channel, $I_m$ be the iteration limit. Table I gives the parameters used in the simulation experiments.

TABLE I
PARAMETERS USED IN THE SIMULATION EXPERIMENTS

| | | | |
|---|---|---|---|
| $CW_{\min}$ | 64 | $a$ | 0.5 |
| $E[L](bits)$ | 2048 | $P_e$ | 0.1 |
| SIFS($\mu s$) | 28 | $m$ | 15 |
| DIFS($\mu s$) | 54 | $c_1$ | 1.5 |
| ACK($bits$) | 240 | $c_2$ | 1.5 |
| $s(\mu s)$ | 13 | $w$ | 0.8 |
| $M$ | 5 | $\triangle cw_{\max}$ | 10 |
| $R(Mbps)$ | 6 | $I_m$ | 300 |

Figure 4 shows the relationship between the end-to-end delay from the first backbone vehicle to the last backbone vehicle and the number of backbone vehicles in a multi-platoon when the standard minimum contention window size is used. According to [24], a packet should be ensured to be accepted successfully within a maximum delay limit 100$ms$. It is seen that the maximum number of backbone vehicles in a multi-platoon is 24.

The optimal minimum contention window sizes under the different number of backbone vehicles $n$ are given by table II. Since there are 2 backbone vehicles in each platoon and the maximum number of backbone vehicles in a multi-platoon is 24. Here $n$=4,…,24.

In Figure 5-9, the minimum contention window size, one-hop delay, end-to-end delay, one-hop throughput and transmission probability with the proposed swarming approach are discussed respectively under 6 backbone vehicles. Figure 5 shows the comparison between the optimal minimum contention window size of each backbone vehicle and the standard minimum contention window size when the number of backbone vehicle is 6. It is seen that all optimal minimum contention window sizes are smaller than the standard minimum contention window size, thus it can guarantee a relatively low one-hop delay. Moreover, the minimum contention window size of each backbone vehicle is different to get the one-hop delay of each backbone vehicle balanced.

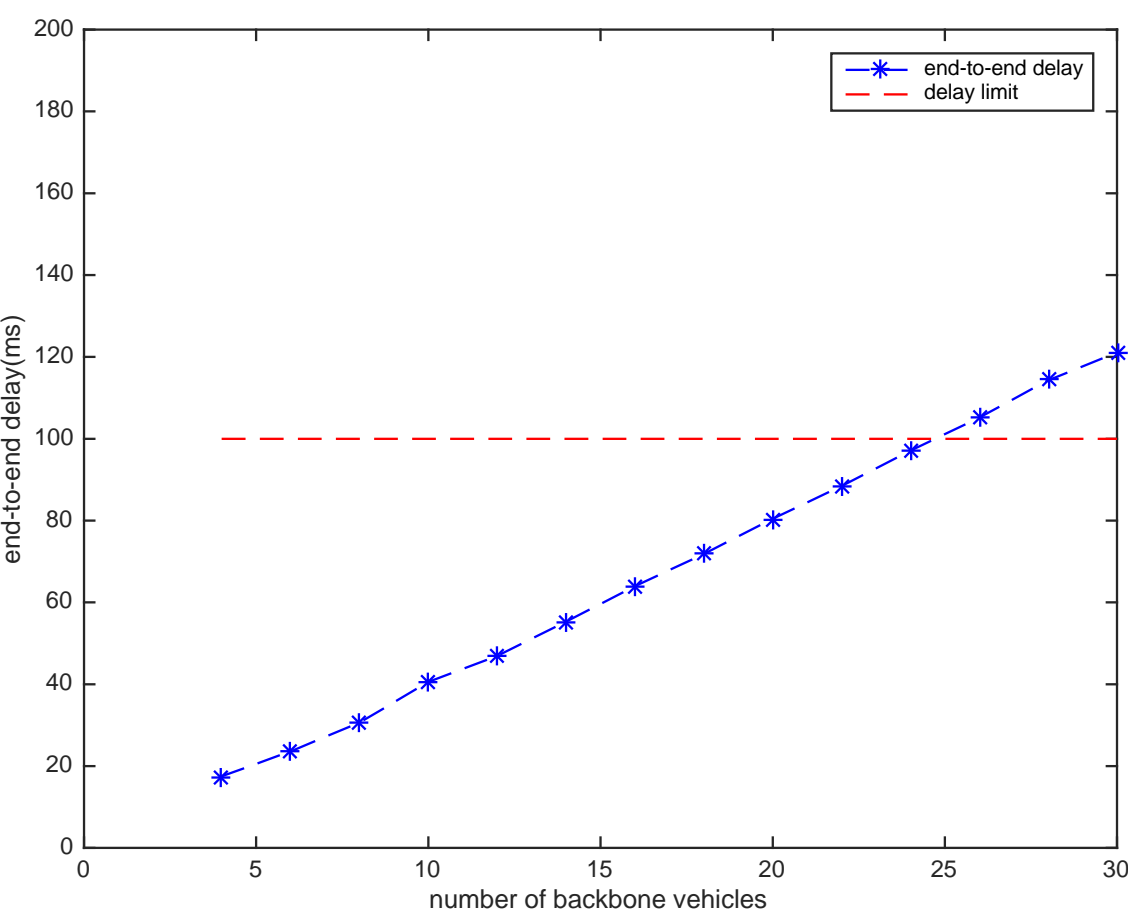

**FIGURE 4. End-to-end delay vs number of backbone vehicles**

TABLE II
THE OPTIMAL CONTENTION WINDOW SIZES

| $n$ | The optimal minimum contention window sizes |
|---|---|
| 4 | [38,49,49,38] |
| 6 | [34,43,20,20,43,34] |
| 8 | [38,55,24,22,22,24,55,38] |
| 10 | [36,51,22,20,18,18,20,22,51,36] |
| 12 | [40,54,22,20,18,18,18,18,20,22,54,40] |
| 14 | [32,45,18,18,17,20,20,20,20,17,18,18,45,32] |
| 16 | [44,56,23,18,16,17,20,21,21,20,17,16,18,23,56,44] |
| 18 | [34,45,17,15,14,15,16,17,18,18,17,16,15,14,15,17,45,34] |
| 20 | [34,50,21,24,23,30,29,30,27,26,26,27,30,29,30,23,24,21,50,34] |
| 22 | [42,55,19,15,13,14,17,19,22,21,21,21,21,22,19,17,14,13,15,19,55,42] |
| 24 | [38,50,20,18,17,20,22,23,27,28,31,32,32,31,28,27,23,22,20,17,18,20, 50,38] |

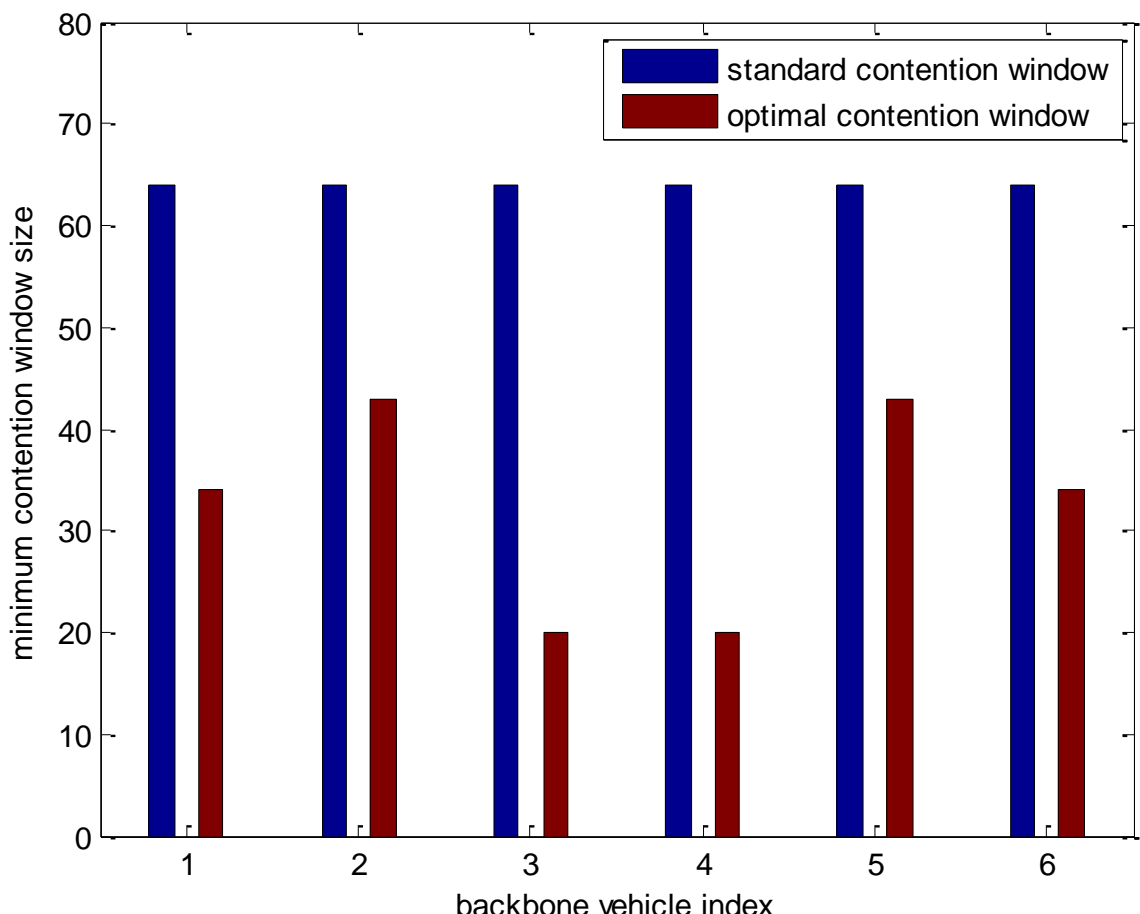

**FIGURE 5. Minimum contention window size vs backbone vehicle index**

Figure 6 shows the comparison between the one-hop delay of each backbone vehicle with the optimal minimum contention window sizes and that with the standard minimum contention window size when the number of backbone vehicles is 6. It is seen that the one-hop delays of backbone vehicles with the standard minimum contention window size are unbalanced. The one-hop delays of backbone vehicles with the optimal minimum contention window size are balanced and kept at around a small value 3.2*ms*.

Figure 7 shows the comparison between the end-to-end delay from backbone vehicle 1 to backbone vehicle $i$ ($i$=2,3,…,6) with the optimal minimum contention window size and that with the standard minimum contention window sizes when the number of backbone vehicles is 6. It is seen that the end-to-end delay from backbone vehicle 1 to backbone vehicle 6 with the optimal minimum contention window sizes is around 16*ms*. This is around 5*ms* lower than that with the standard minimum contention window size. Moreover, the end-to-end delay from backbone vehicle 1 to other backbone vehicles with the optimal minimum contention window sizes is increased by a constant with the increasing of backbone vehicle index. This is because that one-hop delay of each backbone vehicle with the optimal minimum contention window sizes is balanced.

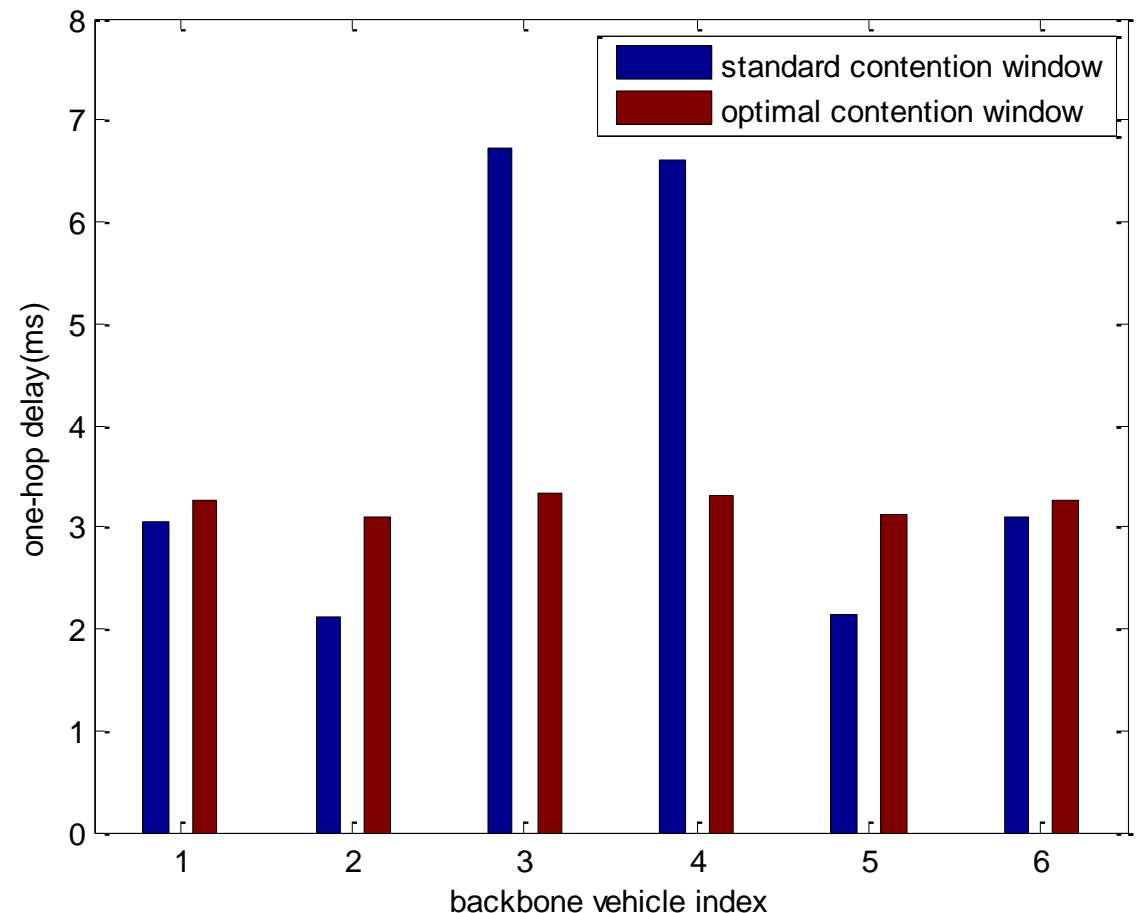

**FIGURE 6. One-hop delay vs backbone vehicle index**

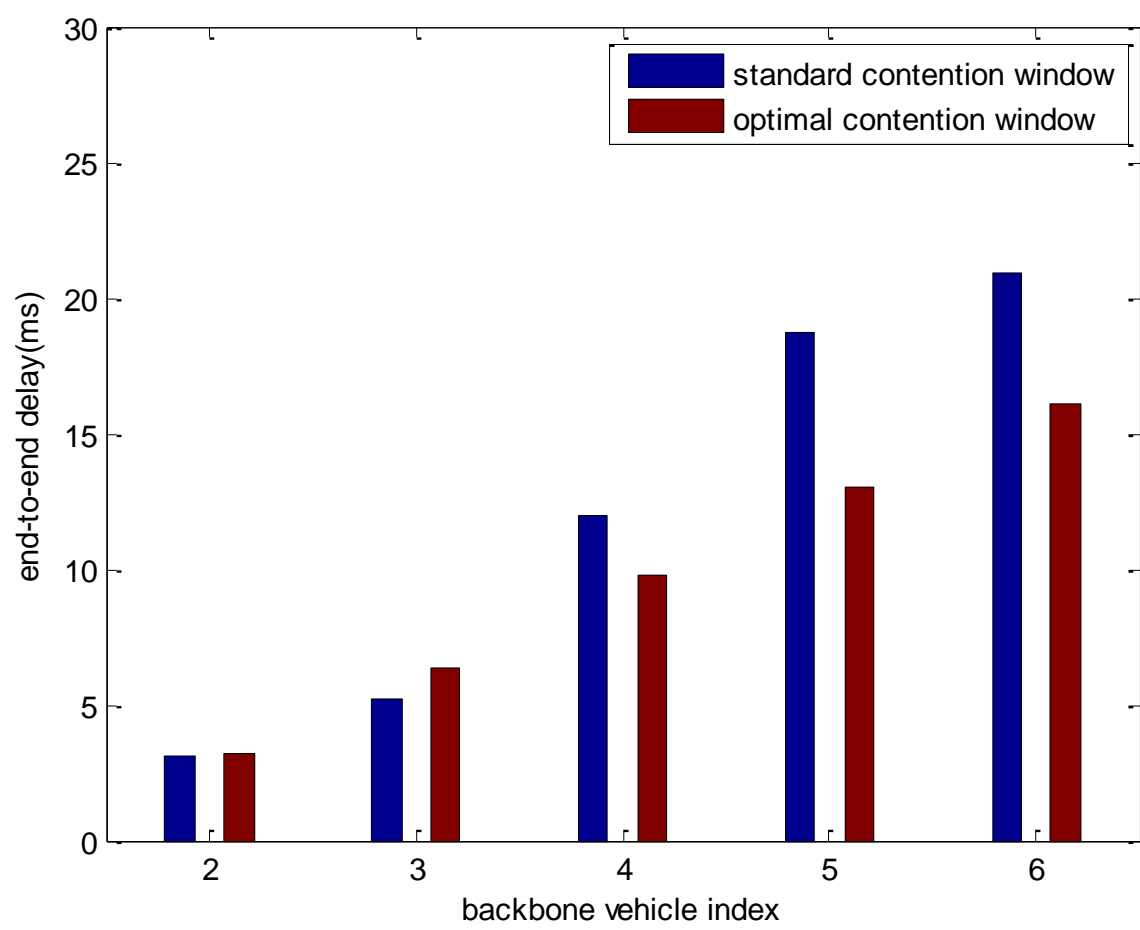

**FIGURE 7. End-to-end delay vs backbone vehicle index**

Figure 8 shows the comparison between the one-hop throughput of each backbone vehicle with the optimal minimum contention window sizes and that with the standard minimum contention window size when the number of backbone vehicles is 6. It is seen that the one-hop throughput of each backbone vehicle with the optimal minimum contention window sizes are kept at around $0.6 Mbp/s$ and the end-to-end throughput is slightly higher than that of each backbone vehicle with the standard minimum contention window.

Figure 9 shows the comparison between the transmission probability of each backbone vehicle with the optimal minimum contention window sizes and that with the standard minimum contention window sizes when the number of backbone vehicles is 6. It is seen that the transmission probability of each backbone vehicle with the optimal minimum contention window sizes is larger than that with the standard minimum contention window size.

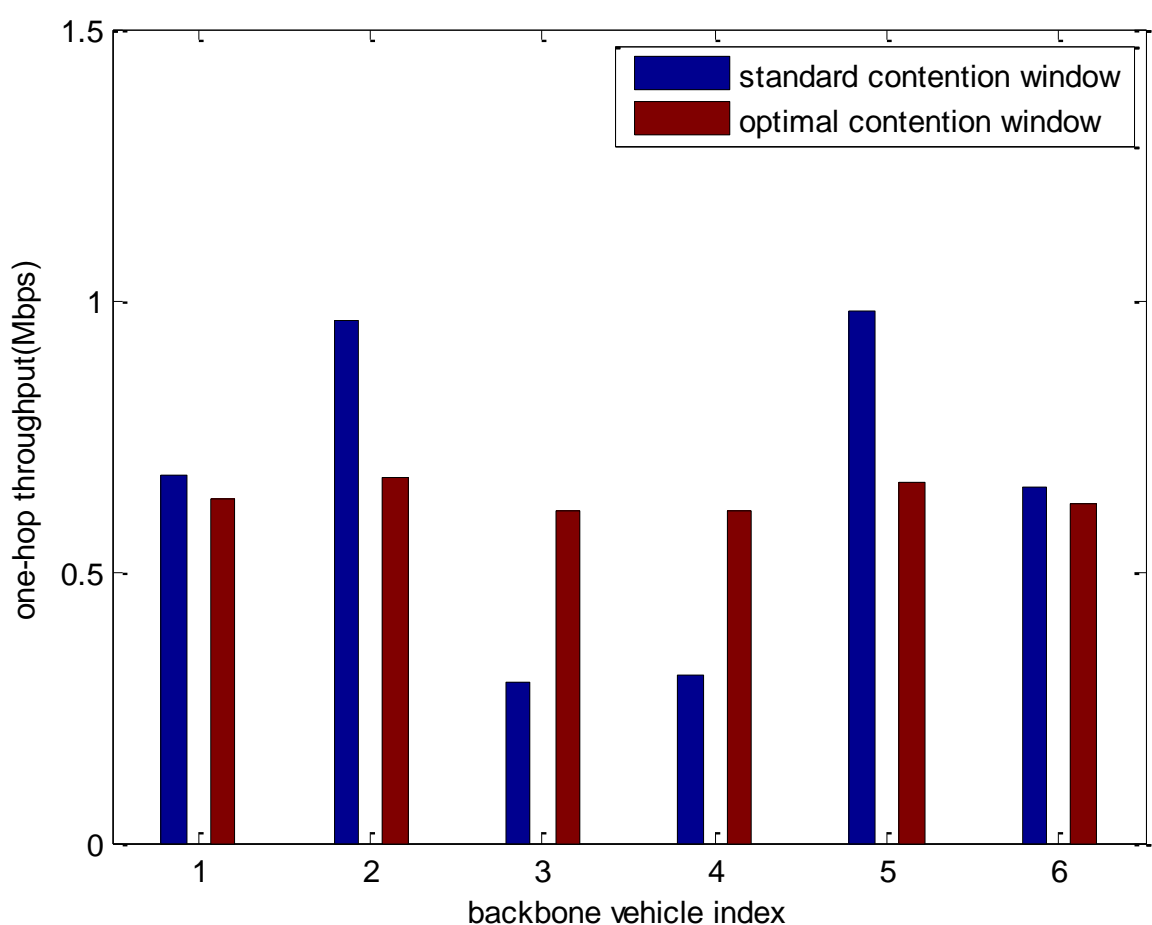

**FIGURE 8. One-hop throughput vs backbone vehicle index**

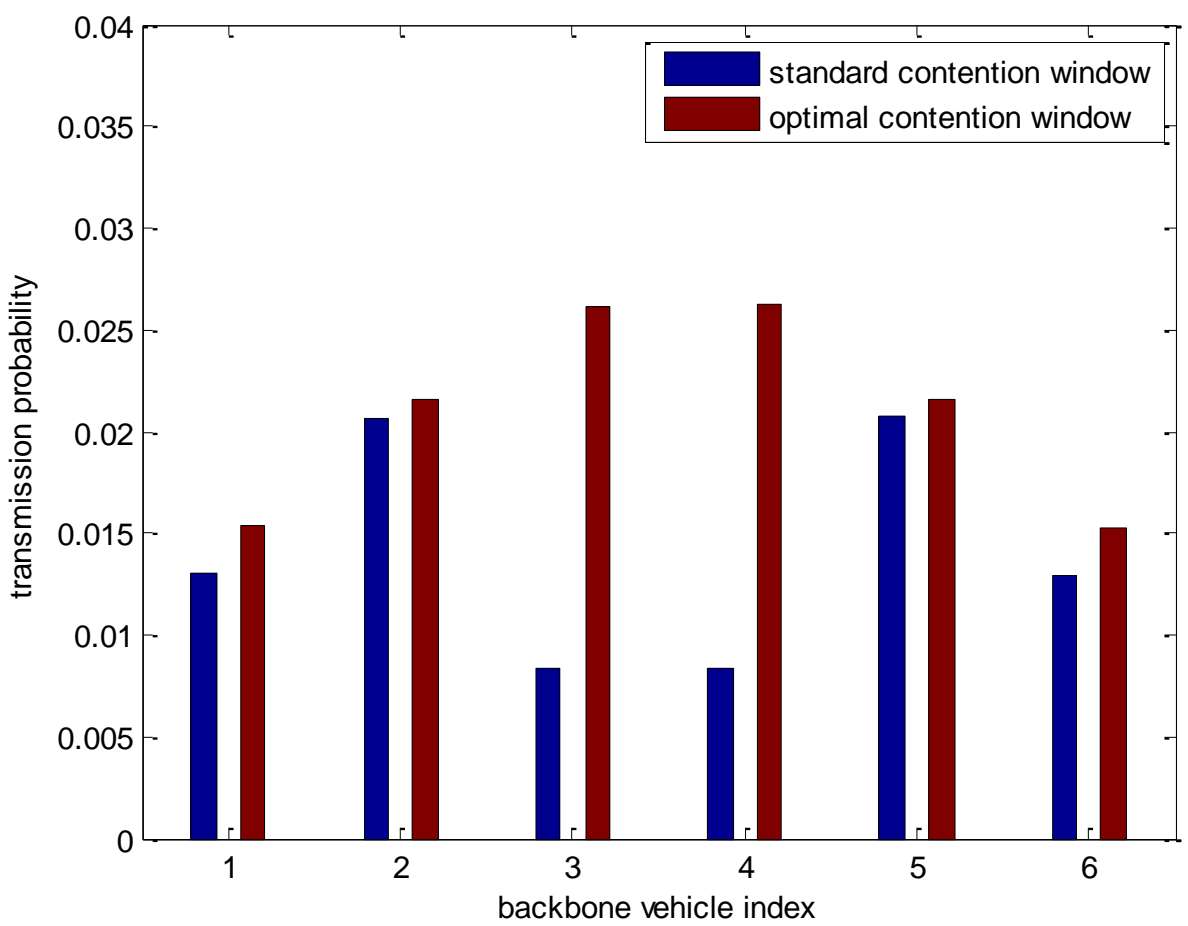

**FIGURE 9. Transmission probability vs backbone vehicle index**

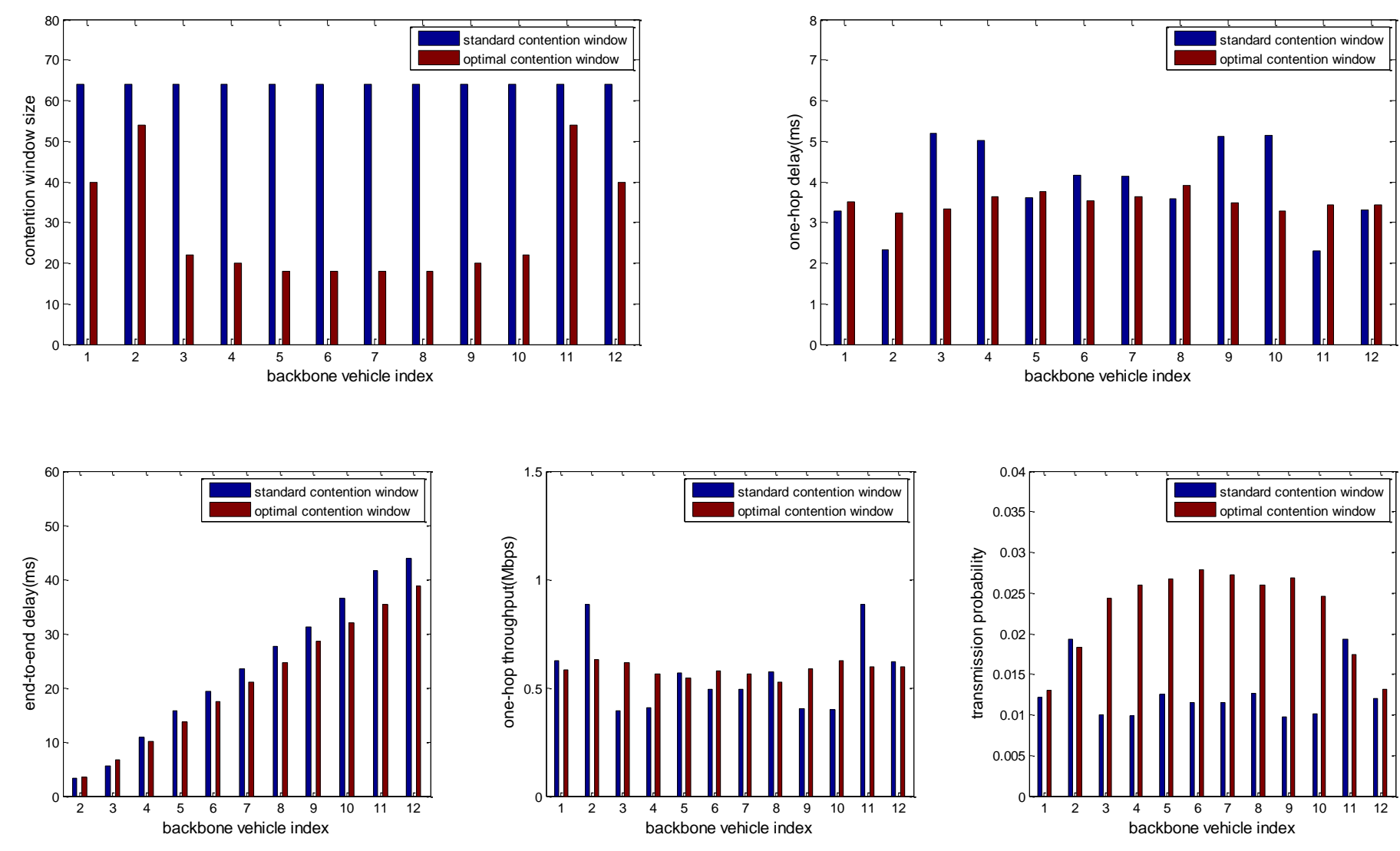

**FIGURE 10. The performance of inter-platoon communications under 12 backbone vehicles**

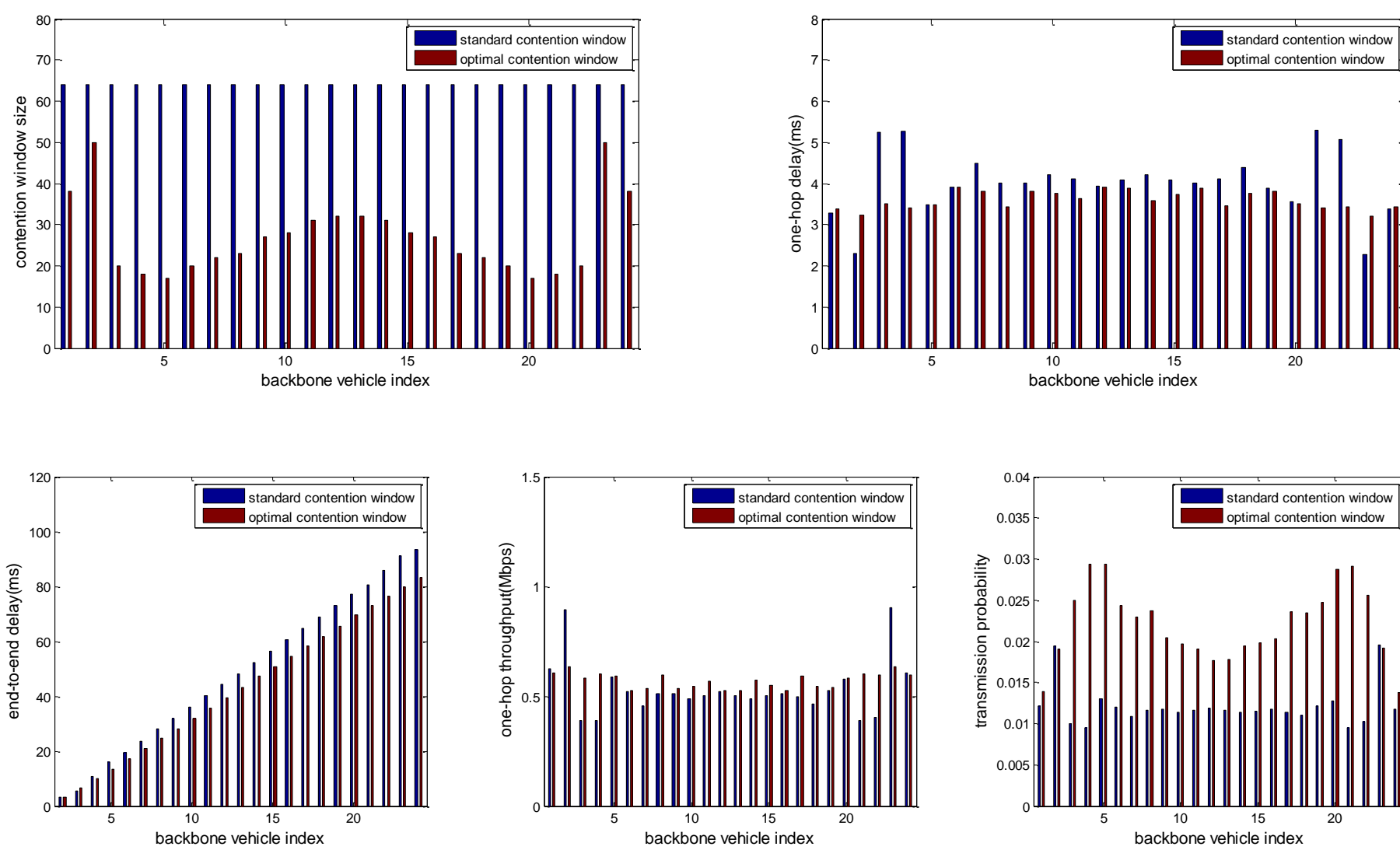

**FIGURE 11. The performance of inter-platoon communications under 24 backbone vehicles**

In Figure 10-11, the performance metrics with the proposed swarming approach are discussed under 12 and 24 backbone vehicles respectively. Figure 10 and Figure 11 compare the performance of inter-platoon communications with the optimal minimum contention window sizes and the standard minimum contention window size when the number of backbone vehicles in a multi-platoon is 12 and 24 respectively. It is seen that the one-hop delay and the one-hop throughput of each backbone vehicle with the optimal minimum contention window sizes are kept at around a constant. The end-to-end delay from backbone vehicle 1 to backbone vehicle $i$ ($i$=2,3,…,12/24) with the optimal minimum contention window sizes is slightly lower than that of each backbone vehicle with the standard minimum contention window. The end-to-end delay with the optimal minimum contention window sizes is slightly lower than that with the standard minimum contention window. The transmission probability of each backbone vehicle with the optimal minimum contention window sizes is larger than that with the standard minimum contention window size.

## VI. CONCLUSION

In this paper, we proposed a swarming approach to optimize the one-hop delay for inter-platoon communications through adjusting the minimum contention window size of each backbone vehicle in two steps. In the first step, we first set a small enough average one-hop delay of backbone vehicles as the initial optimization goal and then proposed a swarming approach to find a minimum average one-hop delay for inter-platoon communications through adjusting the minimum contention window of each backbone vehicle iteratively. In the second step, we first set the minimum average one-hop delay found in the first step as the initial optimization goal and then adopted the swarming approach again to get the one-hop delay of each backbone vehicle balance to the minimum average one-hop delay. The adjusted minimum contention window sizes that get the one-hop delay of each backbone vehicle balance to the minimum average one-hop delay were obtained after the second step. The simulation results indicated that the one-hop delay is optimized and the other performance metrics including end-to-end delay, one-hop throughput and transmission probability are considerably improved by using the optimal minimum contention window sizes, compared with the IEEE standard contention window sizes.